\documentclass[11pt]{article}
\usepackage{epsfig}
\def\Ha{H$\alpha$}
\def\hi{{\sc H\thinspace i}}
\def\hii{{\sc H\thinspace ii}}
\def\Zsol{Z_\odot}
\newcommand{\Lx}{$L_{\rm x}$}
\newcommand{\kms}{\rm km\,s^{-1}}
\newcommand{\ergs}{\rm erg\,s^{-1}}

\newcommand{\etal}{{et\thinspace al.}~}
\def\spose#1{\hbox to 0pt{#1\hss}}
\def\lta{\mathrel{\spose{\lower 3pt\hbox{$\mathchar"218$}}
     \raise 2.0pt\hbox{$\mathchar"13C$}}}
\def\gta{\mathrel{\spose{\lower 3pt\hbox{$\mathchar"218$}}
     \raise 2.0pt\hbox{$\mathchar"13E$}}}
\newcommand{\Dt}{\spose{\raise 1.5ex\hbox{\hskip5pt$\mathchar"201$}}}    
\def\Mdot{\Dt{M}}

\title{{\bf Mechanical Feedback:  \\
	From Stellar Wind Bubbles to Starbursts
}}
\author{M.~S.~Oey$^{1,3}$, C.~J.~Clarke$^2$, 
	P.~Massey$^3$ \\
\vspace{0.1cm}\\
\normalsize $^1$Space Telescope Science Institute, 3700 San Martin Dr.,
	Baltimore, MD   21218, U.S.A.\\
\normalsize $^2$Institute of Astronomy, Madingley Road, 
	Cambridge CB3 0HA, Cambridge, U.K.\\
\normalsize $^3$Lowell Observatory, 1400 W. Mars Hill Rd., Flagstaff,
	AZ   86001, U.S.A.\\
}
\date{}
\begin{document}
\maketitle
\def\bull{\vrule height .9ex width .8ex depth -.1ex}
\makeatletter
\def\ps@plain{\let\@mkboth\gobbletwo
\def\@oddhead{}\def\@oddfoot{\hfil\tiny
``Dwarf Galaxies and their Environment'';
International Conference in Bad Honnef, Germany, 23-27 January 2001}%
\def\@evenhead{}\let\@evenfoot\@oddfoot}
\makeatother

\begin{abstract}\noindent
The current understanding of mechanical feedback is reviewed by
evaluating the standard, adiabatic model for shell formation and
evolution.  This model is relevant to phenomena ranging from individual
stellar-wind bubbles to galactic superwinds, forming the basis
for our understanding of the multiphase ISM, IGM, and galactic
evolutionary processes.  Although significant discrepancies between
the model and observation have been identified, to date there are none
that require a fundamental revision.  A variety of evidence, ranging
over three orders of magnitude in spatial scale, is broadly consistent
with the standard model.  This includes kinematics of individual
objects, observations of hot gas, the size distribution of \hi\
shells, and outflow rates from 
starburst galaxies.  However, some of the most pressing issues
relating to shell evolution are still outstanding and obstruct efforts
to resolve key questions like the fate of the hot gas.
\end{abstract}

\section{Introduction}

Mechanical feedback from massive
stars is one of the principal drivers of evolutionary processes in
galaxies.  Our understanding of the feedback process is grounded in
the conventional model for the evolution of stellar wind- and
supernova-driven superbubbles (e.g., Pikel'ner 1968; Weaver
{\etal}1977):  hot (log $T/{\rm K}\sim6-7$), low-density ($n\sim 0.01\
\rm cm^{-3}$) gas is generated within a double-shock structure, and
the pressure 
of this hot gas, chemically enriched by the stellar products, drives
the growth of the thin, radiatively-cooled shell.  In the standard
model, energy loss from the hot gas is considered to be negligible,
yielding simple analytic expressions for the shell radius $R$ and
expansion velocity $v$ as a function of time $t$:
\begin{eqnarray}\label{eqAD}
R & \propto & (L/n)^{1/5}\ t^{3/5} \quad , \nonumber \\
v = \Dt{R} & \propto & (L/n)^{1/5}\ t^{-2/5} \quad .
\end{eqnarray}
For a stellar wind-driven bubble, the input mechanical luminosity 
$L=1/2\ \Dt{M} v_\infty^2$, where $\Dt{M}$ and $v_\infty$ are the
stellar wind 
mass-loss rate and terminal velocity, respectively.  For OB
associations, supernovae quickly dominate over winds, in which case
$L=N_* E_{51}/t_e$ (e.g., Mac Low \& McCray 1988), where $N_*$ and
$E_{51}$ are the total number of supernovae and supernova energy,
respectively, and $t_e$ is the time over which the supernovae occur.

The fate of this hot gas is crucial in understanding the phase balance
and enrichment of the interstellar and intergalactic media.  Is
stellar feedback indeed the source of the diffuse, hot gas in the ISM?
Do galactic superwinds from starbursts eject gas from galaxies?  How
does mechanical feedback affect the structure of the ISM and porosity,
for example, to ionizing radiation?  Do superbubbles trigger renewed
star formation?  To answer these basic questions, we must first
evaluate the relevance of the standard, adiabatic model for shell
evolution, which is applied over scales
ranging from single stellar wind bubbles to galactic superwinds.

\section{Single star bubbles}

Single star bubbles of isolated OB stars are woefully understudied.
Oey \& Massey (1994) examined two nebular examples in M33, and
spectroscopically classified the late-type O stars.  The inferred
stellar masses and ages implied wind parameters that were consistent
with the observed sizes and shell ages predicted by the adiabatic
model.  However, the parameters were loosely constrained and the shell
kinematics were not observed.  \hi\ shells with radii of several tens
of pc have been identified as wind-blown bubbles around a number of
Galactic O and Of stars (Cappa \& Benaglia 1998; Benaglia \& Cappa
1999).  These are largely consistent with the standard model, and
probe a specific subset of fairly evolved stars with old shells that
have essentially stopped expanding.

Studies of Wolf-Rayet ring nebulae suggest shells that are too small,
equivalent to an overestimate 
in $L/n$ by an order of magnitude (e.g., Treffers \& Chu 1982;
Garc\'\i a-Segura \& Mac Low 1995; Drissen {\etal}1995).  However, the
progenitor star produces several wind phases, including both fast and
slow winds, with strongly changing $\Dt{M}$.  Their cumulative effect 
on shell morphology is even more complex and poorly-understood than
the wind evolution itself, and it is therefore unsurprising to find
significant discrepancies between the predictions and observations of
shell parameters.  Hence it is desireable to study more of the simpler,
single OB star bubbles.

\section{Superbubbles}

Superbubbles created by OB associations are more prominent than
single-star bubbles, and thus have been studied more extensively.
Soft X-ray emission has been detected from the interiors of many
objects in projection, which is qualitatively consistent with the
adiabatic superbubble model.  Quantitatively, two classes of X-ray
emission have been identified:  objects with X-ray luminosity \Lx\ in
excess of the standard model's prediction (Chu \& Mac Low 1990; Wang
\& Helfand 1991), and objects that remain undetected in X-rays (Chu
{\etal}1995).  The X-ray--bright objects are suggested to be
overluminous because of SNR impacts on the shell walls.  Upper limits
on the X-ray--dim objects remain consistent with \Lx\ predicted by the
adiabatic model.  It will thus be of great interest to determine \Lx\
for these objects with {\it XMM-Newton} or {\it Chandra}.  The
existence of hot gas within superbubbles also implies an interface
region with the cooler shells where intermediate temperatures and ions
should be present.  Chu {\etal}(1994) searched a number sightlines
through LMC superbubbles and confirmed the existence of
C\thinspace{\sc iv} and Si\thinspace{\sc iv} absorption in all cases.

\begin{figure*}[tb]
\begin{center}
\epsfig{file=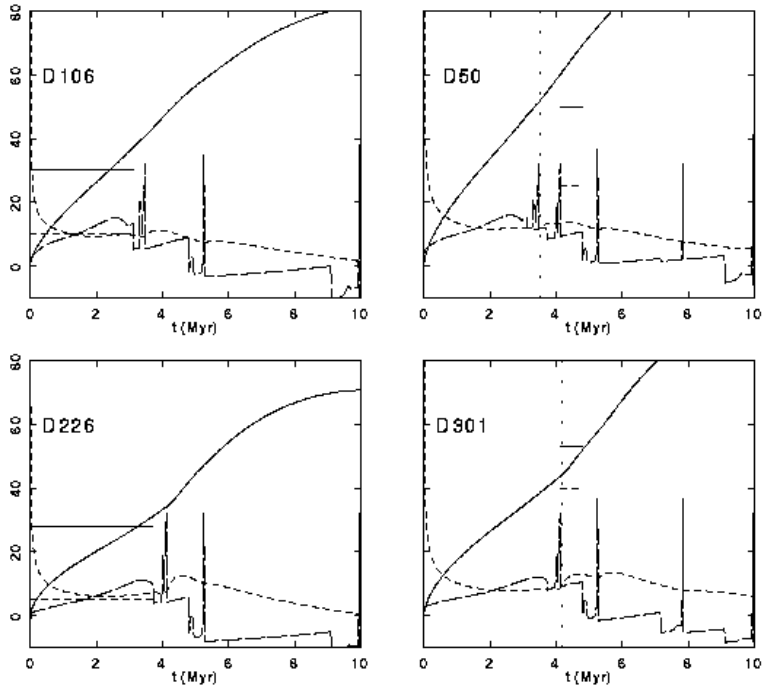,width=11cm}
\caption[]{\small Models for the evolution of LMC superbubbles, with $L/n$
reduced by a factor of 10 from the empirical estimate.  Solid, dotted,
and dashed lines show the predicted $R$ (pc), $v\ (\kms$), and input \break
10 log $L/(10^{35}\ \ergs)$, respectively, as a function of $t$ (Myr).
The horizontal solid and dotted lines show the observed $R$ and $v$,
respectively, with the time constraints set by the observed stellar
population.  
\label{FO96} }
\end{center}
\end{figure*}

A stringent test of the adiabatic model is to compare the predicted
and observed shell kinematics where the input mechanical power and
other parameters are well-constrained.  We obtained spectroscopic
classifications of the OB associations in eight LMC superbubbles 
(Oey \& Massey 1995; Oey 1996; Oey \& Smedley 1998),
thereby providing estimates of $L$ in these young, wind-dominated
objects.  The predicted growth rate for the
shells was higher than implied by their observed $R$ and $v$,
equivalent to an overestimate in $L/n$ by 
an order of magnitude.  However, even after adjusting $L/n$ in the
models, over half the objects still showed observed expansion
velocities that were typically a factor of two higher than predicted
for the given $R$.  Similar discrepancies were reported for
Galactic objects by Saken {\etal}(1992) and Brown {\etal}(1995).
Figure~\ref{FO96} shows representative
examples:  the two objects in the left column show predicted $R$
(solid curves) and $v$ (dotted curves) in agreement with the observed
values (horizontal segments); while the objects on the right show
anomalously high observed $v$.  All the models have $L/n$ reduced by a
factor of 10.  The objects with anomalously high $v$ also each exhibit
anomalously high 
\Lx\ and [S\thinspace{\sc ii}]/\Ha, consistent with the suggestion by
Chu \& Mac Low (1990) and Wang \& Helfand (1991) that SNR impacts have
heated the shell.  These would then also accelerate the shell walls,
explaining the high observed $v$.  The observed stellar mass function
in these objects also support this scenario, implying the former
existence of 1--2 higher-mass stars corresponding to SN
progenitors. 


While the SNR impact hypothesis appears the most likely explanation 
for the high-velocity shells, a sudden drop in the ambient density can
induce a ``mini-blowout'' with shell kinematics that can easily
reproduce the anomalous observations (Oey \& Smedley 1998; Mac Low
{\etal}1998; Franco \& Silich 1999).  Indeed, were it not for the high
X-ray emission and additional evidence above, it would be impossible
to distinguish between the two shell acceleration mechanisms from the
kinematics alone. 

We therefore see that the ambient density distribution is critical in
determining the shell evolution.  An underestimate in $n$ could, for
example, contribute to the growth rate discrepancy described above,
that is seen in all the objects.  To clarify the effect of the ambient
gas distribution, Oey {\etal}(2001) mapped the \hi\ distribution
within a $\sim40^\prime$ radius of three nebular LMC superbubbles at
30$^{\prime\prime}$ resolution, using the Australia Telescope Compact
Array.  The results show neutral environments 
that vary to an extreme:  DEM L25 shows no \hi\ shell component, but
appears to be nestled in an \hi\ hole; DEM L50 shows an \hi\ shell
with $\sim10$ times more mass than the \Ha\ component, but
is otherwise in a large (1.4 kpc) \hi\ void; and DEM L301
is in a non-descript neutral environment with no apparent
correspondence to the optical shell.  Thus, despite the similar
appearance of the \Ha\ nebulae, the \hi\ distributions could not be
more different.  It is therefore extremely difficult to infer
properties of ambient density distribution without direct, empirical
determinations.

Another vital parameter for shell evolution is the ambient pressure,
$P_0$, which determines whether and when the superbubble growth becomes
pressure-confined.  While $P_0$ is usually relatively unimportant in young,
high-pressure superbubbles such as the nebular objects modeled above,
it is of vital importance in the mid- to late-stage evolution.  It may
also be relevant in high-pressure, ionized environments like dense
star-forming regions (e.g., Garc\'\i a-Segura \& Franco 1996).  The
value of $P_0$ ultimately determines the final size of the objects,
and conditions relative to blowout.  The uniformity and distribution
of $P_0$ in the multiphase ISM is therefore especially relevant to the
global effect of superbubbles on the ISM (see below). 

Finally, if the hot gas within superbubbles does not somehow blow out
and merge into the hot, ionized medium (HIM), it
is likely that the objects will cool and depart from energy conservation.
Indeed, cooling of the hot interior has long been one of the principal
questions for superbubble evolution and the fate of the hot gas.
Thermal conduction at the interface between the cool shell wall and
hot gas should cause a high rate of mass-loading into the interior,
by evaporation.  The evaporated shell material strongly dominates the
mass in the hot region, which could be further supplemented by
evaporation and ablation from small clouds overrun by the expanding
shocks (e.g., Cowie \& McKee 1977; McKee {\etal}1984; Arthur \&
Henney 1996).  If the interior density is 
sufficiently increased, radiative cooling will dominate, and the
shells will no longer grow adiabatically.  In addition, Silich
{\etal}(2001) point out the importance of enhanced metallicity
in the superbubble interiors, caused by the stellar and SN yields.
Preliminary investigation for individual objects by Oey \& Silich
(2001, in preparation) shows enhancement in \Lx\ by almost an order of
magnitude for low-metallicity ($Z=0.05\ \Zsol$) objects.  The
elemental yields therefore may also significantly enhance the cooling
rate, facilitating the transition from adiabatic to
momentum-conserving evolution. 

\section{Global mechanical feedback and superwinds}

Quantifying the statistical properties of superbubbles in galaxies
allows us to test the evolutionary model and also gain insight on the
properties of the ISM.  In particular, the so-called interstellar
porosity (see below) is a conventional parameter for evaluating the
relative importance of the HIM in the multiphase ISM.  

\subsection{Superbubble populations}

Oey \& Clarke (1997) derived expressions for the differential size
distribution $N(R)\ dR$ of superbubbles in a uniform ISM, using the
analytic expressions for adiabatic evolution (equation~\ref{eqAD}).
We considered a power-law mechanical luminosity function for the
parent OB associations, 
\begin{equation}\label{eqMLF}
\phi(L)\ dL = A L^{-\beta}\ dL \quad, 
\end{equation}
with $\beta=2$, based on the typical \hii\ region luminosity function
(\hii\ LF; e.g., Kennicutt {\etal}1989; Oey \& Clarke 1998a).
The superbubble growth was also taken to be pressure-confined when the
interior pressure $P_i = P_0$.  Star formation is assumed to be coeval
within each OB association, with SNe therefore exploding over a period
$t_e = 40$ Myr, the lifetime of the lowest-mass SN progenitors.  These
simple assumptions yield two evolutionary paths for the superbubbles:
objects whose growth stalls at radius $R_{\rm f}$ due to pressure
confinement at time $t_{\rm f} \propto (n/P_0)^{1/2}\ R_{\rm f}$; and
objects whose growth continues until $t_e$.  The characteristic
timescale $t_e$ therefore determines the characteristic parameters
$R_e$ and $L_e$, which correspond to conditions for which an object
stalls at $t_e$.

The resultant size distribution for a constant superbubble creation
rate $\psi$, and single-valued $\phi(L)$ is:
\begin{equation}\label{eqcaseA}
N_{\rm grow}(R) \propto R^{2/3} \quad .
\end{equation}
This is determined exclusively by growing objects, since the stall
radius $R_{\rm f}$ in this case is the same for all the shells, and
represents the maximum $R$ in the distribution.

If all objects are created in a single burst, but now with a power-law
$\phi(L)$ (equation~\ref{eqMLF}), then both growing and
stalled objects contribute to the size distribution:
\begin{eqnarray}
N_{\rm grow}(R) & \propto & R^{4-5\beta}\propto R^{-3}  \nonumber \\
N_{\rm stall}(R) & \propto & R^{1-2\beta}\propto R^{-6} \quad ,
\end{eqnarray}
with $\beta=2$.  The stalled objects dominate the total $N(R)$ for
$R\leq R_{\rm f}(t_e)$, beyond which the growing objects dominate, falling
off with a much steeper $R$-dependence.  

For constant $\psi$ and power-law $\phi(L)$, the stalled
objects again dominate, with a total $N(R)$:
\begin{equation}\label{eqcaseC}
N(R) = A\psi\frac{L_e^{1-\beta}}{R_e}\
	\biggl(\frac{R}{R_e}\biggr)^{1-2\beta}\ 
	\biggl[2(t_e+t_s) + \frac{9-6\beta}{-2+3\beta} t_e\ 
		\biggl(\frac{R}{R_e}\biggr)\biggr] \quad ,
\end{equation}
where $t_s$ is an arbitrary period after $t_e$ during the which the
superbubbles survive before disappearing.  Here again, the stalled
objects dominate, effectively yielding $N(R) \propto R^{1-2\beta}
\propto R^{-3}$.  The maximum in $N(R)$ is at $R_{\rm f}(L_{\rm
min})$, corresponding to single-progenitor SNRs.  At $R\leq R_{\rm
f}(L_{\rm min})$, growing, high-$L$ objects dominate with $N(R)
\propto R^{2/3}$ (equation~\ref{eqcaseA}).  Oey \& Clarke (1998b) show
schematic representations of $N(R)$ for the above three combinations of
$\psi$ and $\phi(L)$.

We compared this result to the \hi\ shell catalog for the Small
Magellanic Cloud (SMC) compiled by Staveley-Smith {\etal}(1997).  This
is by far the most complete sample of \hi\ shells obtained for any
galaxy, as evidenced by the fact that the relative number counts of
\hii\ regions and \hi\ shells are in excellent agreement with their
relative life expectancies.  For shells having $R\geq100$ pc, the
fitted power-law slope $\alpha=1-2\beta$ is $2.7\pm 0.6$, in excellent
agreement with the general prediction of $\alpha=3$.  Using
$\beta=1.9$ from the observed \hii\ LF for the
SMC (Kennicutt {\etal}1989) yields a specific SMC prediction of
$\alpha=2.8\pm 0.4$, still in remarkable agreement with the observed
value (Oey \& Clarke 1997).  

We note that different models for ISM structure
yield different predictions for $N(R)$.  For example, Stanimirovi\'c
{\etal}(1999) suggest a possible fractal structure for the neutral ISM.  From
the same \hi\ dataset of the SMC, they find a fractal dimension implying
a size distribution for \hi\ holes of $\alpha=3.5$.  Within
the uncertainties, this prediction cannot be empirically differentiated
from our model with $\alpha=3$; but it is worth noting that
the predictions are in fact intrinsically different.  

However, the superbubble size distribution presently is not a
sensitive test in determining whether or not the objects evolve
adiabatically.  If all the internal energy is radiated away,
the objects are predicted to follow the momentum-conserving law given
by Steigman {\etal}(1975):
\begin{equation}
R\propto (L/nv_\infty)^{1/4}\ t^{1/2} \quad .
\end{equation}
The stall radius $R_{\rm f}$ in this case is only
1.3 times larger than for the adiabatic model.  Furthermore, the size
distribution follows the same law, $N(R)\propto R^{1-2\beta}$ (Oey \&
Clarke 1997).  The observations of hot gas are therefore vital
confirmation that the adiabatic model applies to a significant
fraction of superbubbles. 

It is also possible to derive the distribution of expansion
velocities $N(v)\ dv$.  This is naturally determined only by the
growing objects (Oey \& Clarke 1998b):
\begin{equation}
N_{\rm grow}(v) \propto v^{-7/2}\ , \quad \beta > 1.5 \quad .
\end{equation}
We again compared this prediction to the SMC \hi\ shell catalog, and
found a fitted power-law slope of $2.9\pm 1.4$, which is also in agreement
with the prediction.  Thus, in spite of the crude assumptions in
deriving the shell size and velocity distributions, the data
suggest that the neutral ISM in the SMC is fully consistent
with superbubble activity dominating the structure.  Although most
other available \hi\ shell catalogs are highly incomplete,
preliminary results for a few other galaxies also show agreement with
our model for the size distribution (Kim {\etal}1999; Mashchenko
{\etal}1999; Oey \& Clarke 1997).

\subsection{ISM porosity and galactic superwinds}

It is straightforward to use the analytic expression for $N(R)$
(equation~\ref{eqcaseC}) to derive the interstellar porosity parameter
$Q$, which is given by Oey \& Clarke (1997) for 2- and 3-dimensional
cases.  $Q$ is the ratio of:  total area or volume occupied by
superbubbles, to the total area or volume of the galaxy.  Thus it is
essentially the filling factor of hot gas, assuming hot gas is contained
within all of the superbubbles.  Determining $Q$ has been the
conventional way to evaluate the relative importance of the HIM (e.g.,
Cox \& Smith 1974; McKee \& Ostriker 1977; Heiles 1990).  Values
near unity indicate the HIM dominates the multiphase ISM, and values
significantly $> 1$ imply an outflow, with the galaxy generating more
hot gas than it can contain.

We can write $Q$ in terms of a galaxy's star-formation rate (SFR), $\Psi$:
\begin{equation}\label{eqQSFR}
Q \simeq 16 \frac{\Psi({\rm M_\odot\ yr^{-1}})}{hR_g^2(\rm kpc^3)}
	\qquad \propto \frac{1}{P_0} \ ,
\end{equation}
for $\beta=2$, a Salpeter (1955) IMF for stellar masses $0.1 \leq m
\leq 100\ \rm M_\odot$, and $P_0/k = 9500$.  $R_g$ and $h$ are the
galaxy's gaseous radius and scale height, respectively.  We caution
that $Q$ depends on ambient interstellar parameters, for example,
$P_0^{-1}$ as indicated. 

Table~1 shows the interstellar porosities for all the galaxies in the
Local Group.  Dwarf galaxy parameters are taken from Mateo (1998), and
SFR's for M31, M33, and the Magellanic Clouds from their \hii\ LF's
(Kennicutt {\etal}1989).  In general,
we see that almost all the galaxies show low values with $Q < 0.5$,
implying that the HIM is not the dominant ISM component in these
cases.  The LMC is an exception, showing $Q\sim 1$, so the HIM
appears likely to dominate in this galaxy.  The Milky Way results
historically have been ambiguous and controversial (e.g., McKee \&
Ostriker 1977; Slavin \& Cox 1993), and we obtain different answers
from different methods.  Taking the Galaxy's SFR from
the \hii\ LF of McKee \& Williams (1997) and Smith \& Kennicutt (1989)
yields a low $Q\sim 0.2$; but using the SN rate of van den Bergh \&
Tammann (1991) yields $Q\gta 1$.

\begin{table}
\centerline{Table 1:  Porosities in the Local Group}
\begin{center}
\begin{tabular}{ll}
\\
Galaxy & Q \\
\hline
M31	& 0.03 \\
M33	& 0.3  \\
LMC	& 1.0  \\
SMC	& 0.3  \\
IC~10	& 23   \\
LG dwarfs & 0.01 -- 0.2 \\
\\
Milky Way (\hii\ LF) & 0.2 \\
Milky Way (SN rate)  & $\sim$1 \\
\hline
\end{tabular}
\end{center}
\end{table}

However, by far the highest porosity is found for IC~10, the one
starburst galaxy in the Local Group.  IC~10 shows a $Q$ value that is
over an order of magnitude larger than for any other galaxy examined,
therefore fulfilling the breakout criterion for a galactic
superwind outflow.  It is possible to estimate the mass-loss rate in
this outflow $\Mdot_{\rm out}$, assuming that the material is largely
evaporated from shell walls by thermal conduction.  Mac Low \& McCray
(1988) have estimated the evaporation rate for an individual
superbubble:
\begin{equation}
\Mdot(R)\simeq 16\pi\mu m_{\rm H}/25k\ CT^{5/2}\ R \quad ,
\end{equation}
where $\mu m_{\rm H}$ is the mean mass per nucleon and $C$ is a
conductivity coefficient.  Integrating this over the superbubble size
distribution yields $\Mdot_{\rm out}$:
\begin{equation}
\Mdot_{\rm out} \simeq \int^{R_{\rm max}}_{R_{\rm min}} \Mdot(R)\ N(R)\
	dR \quad .
\end{equation}
The estimate of $\Mdot_{\rm out}$ is necessarily crude, for example,
we rather arbitrarily take $R_{\rm max} = R_e$, with the caveat that
it may be limited by $R_g$.  

The maximum $\Mdot_{\rm out}$ is obtained for the largest evaporating
surface area of superbubbles, which occurs for $Q\sim 1$.  Porosities
of $Q>1$ imply strong interaction and destruction of the shells, or
blowouts, probably reducing the available surface area.  For IC~10, the
derived normalization for $N(R)$ yielding $Q = 1$ corresponds to
$\Mdot_{\rm out}\sim 9\times 10^{-3}\ \rm M_\odot\ yr^{-1}$.  For
comparison, the SFR required for $Q=1$ (equation~\ref{eqQSFR}) is
$\Psi \sim 3\times 10^{-2}\ \rm M_\odot\ yr^{-1}$, which is only about
3 times more than $\Mdot_{\rm out}$!  Although the actual SFR for
IC~10 is much larger ($7\times 10^{-1}\  \rm M_\odot\ yr^{-1}$), yielding
the higher original $Q$ value, this rough calculation suggests that
$\Mdot_{\rm out}$ from galactic winds can be of similar magnitude as
the parent starburst SFR.  Indeed, an extensive absorption-line study
of local starburst galaxies by Heckman {\etal}(2000) also concluded
that empirically, the outflow and star-formation rates have the same
order of magnitude for their sample.

Finally, we note the extensive body of numerical work on superbubbles and
blowout conditions.  These are too numerous to discuss here, but we
refer the reader to Mac Low (1999) and Strickland \& Stevens (2000),
who provide overviews of this field.  The details of the numerical
predictions are presently difficult to confirm empirically, but
observations with {\it XMM-Newton} and {\it Chandra} will be
especially helpful to constrain the dominant processes. 

\section{Summary}

We have seen that observations of mechanical feedback ranging from
individual stellar wind bubbles to galactic superwinds are all
roughly consistent with the conventional adiabatic model for bubble
evolution.  The empirical confirmation of hot, coronal gas
supports the relevance of the adiabatic evolution for at least an
important proportion of shell phenomena.  Although kinematic
discrepancies are often found for individual objects, these anomalies
have plausible explanations, and presently there are no cases that suggest a
need for any major revision of the conventional understanding.  It is a
remarkable strength that the adiabatic model apparently applies to
phenomena over a range of size scales covering at least three orders
of magnitude, from pc to kpc. 

However, characterizing the dominant parameters and their effects on
the shell evolution is still highly problematic.  For example,
critical ambient ISM conditions like the density, pressure, and ionization
distributions remain elusive.  The mechanisms and conditions for
cooling of the interior energy need to be identified, and the
energy budgets reliably determined.  Perhaps the most
fundamental question is the fate of the hot gas generated within the
superbubbles:  Does it escape and thereby constitute the HIM?  Does
it indeed escape from starburst galaxies, and from their gravitational
potentials?  These issues have vital consequences for almost all
galactic evolutionary processes, and our understanding depends on
further clarifying the bubble and superbubble process.

\bigskip
\thanks
MSO gratefully acknowledges travel support from the Graduiertenkolleg
and STScI Institute Fellowship.  Some material in \S 4.2 was derived
in 1998 at the Aspen Center for Physics.

{\small
\begin{description}{} \itemsep=0pt \parsep=0pt \parskip=0pt \labelsep=0pt
\item {\bf References}

\item
\item Arthur, S. J. \& Henney, W. J. 1996, ApJ, 457, 752
\item Benaglia, P. \& Cappa, C. E. 1999, A\&A, 346, 979
\item Brown, A. G. A., Hartmann, D., \& Burton, W. B. 1995, A\&A, 300, 903
\item Cappa, C. E. \& Benaglia, P. 1998, AJ, 116, 1906
\item Chu, Y.-H., Chang, H.-W., Su, Y.-L., \& Mac Low, M.-M.,
	1995, ApJ, 450, 157
\item Chu, Y.-H. \& Mac Low, M-M., 1990, ApJ, 365, 510
\item Chu, Y.-H., Wakker, B., Mac Low, M.-M., \& Garc\'\i a-Segura, G.
	1994, AJ, 108, 1696
\item Cowie, L. L. \& McKee, C. F. 1977, ApJ, 211, 135
\item Cox, D. P. \& Smith, B. W. 1974, ApJ, 189, L105
\item Drissen, L., Moffat, A. F. J., Walborn, N. R., \& Shara,
	M. R. 1995, AJ, 110, 2235
\item Garc\'\i a-Segura, G. \& Franco, J. 1996, ApJ, 469, 171
\item Garc\'\i a-Segura, G. \& Mac Low, M.-M. 1995, ApJ, 455, 145
\item Heckman, T. M., Lehnert, M. D., Strickland, D. K., \& Armus,
	L. 2000, ApJS, 129, 493
\item Heiles, C. 1990, ApJ, 354, 483
\item Kennicutt, R. C., Edgar, B. K., \& Hodge, P. W. 1989, ApJ, 337, 761
\item Kim, S., Dopita, M. A., Staveley-Smith, L., \& Bessell,
	M. S. 1999, AJ, 118, 2797
\item Mac Low, M.-M., in {\sl New Perspectives on the Interstellar
Medium,} eds. A. R. Taylor, T. L. Landecker, \& G. Joncas, San
	Francisco:  Astron. Soc. Pacific, 303
\item Mac Low, M.-M., Chang, T. H., Chu, Y.-H., Points, S. D., Smith,
	R. C., \& Wakker, B. P. 1998, ApJ, 493, 260
\item Mac Low, M.-M. \& McCray, R. 1988, ApJ, 324, 776
\item Magnier, E. A., Chu, Y.-H., Points, S. D., Hwang, U., \&
	Smith, R. C. 1996, ApJ, 464, 829
\item Mashchenko, S. Y., Thilker, D. A., \& Braun, R. 1999, A\&A, 343,
	352
\item Mateo, M. L. 1998, ARAA, 36, 435
\item McKee, C. F. \& Ostriker, J. P. 1977, ApJ, 218, 148
\item McKee, C. F., Van Buren, D., \& Lazareff, B. 1984, ApJ, 278, L115
\item McKee, C. F. \& Williams, J. P. 1997, ApJ, 476, 144
\item Oey, M. S., 1996, ApJ, 467, 666
\item Oey, M. S. \& Clarke, C. J. 1997, MNRAS, 289, 570
\item Oey, M. S. \& Clarke, C. J. 1998a, AJ, 115, 1543
\item Oey, M. S. \& Clarke, C. J. 1998b, in {\sl Interstellar
	Turbulence,} eds. J. Franco \& A. Carrami\~nana, Cambridge:
	Cambridge Univ. Press, 112
\item Oey, M. S., Groves, B., Staveley-Smith, L., \& Smith,
	R. C. 2001, in preparation
\item Oey, M. S. \& Massey, P., 1994, ApJ, 425, 635
\item Oey, M. S. \& Massey, P., 1995, ApJ, 452, 210
\item Oey, M. S. \& Smedley, S. A. 1998, AJ, 116, 1263
\item Pikel'ner, S. B. 1968, Astrophys. Lett., 2, 97
\item Saken, J. M., Shull, J. M., Garmany, C. D., Nichols-Bohlin, J.,
	\& Fesen, R. A. 1992, ApJ, 397, 537
\item Salpeter, E. E. 1955, ApJ, 121, 161
\item Silich, S. \& Franco, J. 1999, ApJ, 522, 863
\item Silich, S. A., Tenorio-Tagle, G., Terlevich, R., Terlevich, E.,
	\& Netzer, H. 2001, MNRAS in press, astro-ph/0009092
\item Slavin, J. D. \& Cox, D. P. 1993, ApJ, 417, 187
\item Smith, T. R. \& Kennicutt, R. C. 1989, PASP, 101, 649
\item Stanimirovi\'c, S., Staveley-Smith, L., Dickey, J. M., Sault,
	R. J., \& Snowden, S. L. 1999, MNRAS, 302, 417
\item Staveley-Smith, L., Sault, R. J., Hatzidimitriou, D., Kesteven,
	M. J., \& McConnell, D. 1997, MNRAS 289, 225
\item Steigman, G., Strittmatter, P. A., \& Williams, R. E. 1975, ApJ,
	198, 575
\item Strickland, D. K. \& Stevens, I. R. 2000, MNRAS, 314, 511
\item Treffers, R. R. \& Chu, Y.-H. 1982, ApJ, 254, 569
\item van den Bergh, S. \& Tammann, G. A. 1991, ARAA, 29, 363
\item Wang, Q. \& Helfand, D. J. 1991, ApJ, 373, 497
\item Weaver, R., McCray, R., Castor, J., Shapiro, P., \& Moore,
	R. 1977, ApJ, 218, 377

\end{description}
}

\end{document}